\documentclass{article}
\usepackage[utf8]{inputenc}
\usepackage{amsmath}
\usepackage{amssymb}
\usepackage{graphicx}
\usepackage{epstopdf}
\usepackage{pgfplots}
\usepackage{booktabs}
\usepackage{makecell}

\pdfoutput=1

\def\beq{\begin{equation}}
\def\eeq{\end{equation}}
\def\bea{\begin{eqnarray}}
\def\eea{\end{eqnarray}}
\def\ar{\begin{array}}
\def\ear{\end{array}}

\def\nn{\nonumber}

\def\ga{\gamma}

\def\de{\delta}

\def\eps{\epsilon}

\begin{document}
\begin{center}
{\Large Ring Type Structures in the Planck map of the CMB}\\[5mm]

Daniel An$^1$, Krzysztof A. Meissner$^{2}$ and Pawe\l~ Nurowski$^{3}$\\[3mm]
{\it $^1$Science Department SUNY Maritime College, 6 Pennyfield Av., New York  10465, USA\\
$^2$ Faculty of Physics, University of Warsaw, Pasteura 5, 02-093 Warsaw, Poland\\
$^3$ Center for Theoretical Physics of PAS, Al. Lotnik{\'o}w 32/46, 02-688 Warsaw, Poland
}

\vspace{3mm}
\begin{abstract}
\noindent We present the results of the quest for ring-type structures on the maps observed by the Planck satellite. The results show that the vicinity of one radius ($\ga=0.14$ radians) of the rings is distinguished. Twisting the circles into deformed ellipses gives a pronounced drop of significance with the increase of twisting.
\end{abstract}
\end{center}

\vspace{0.4cm}

\noindent

\section{Introduction}

The earliest electro-magnetic image of the Universe we have is the cosmic microwave background -- the earlier state of plasma precludes reaching further back.  At almost the same time as the last recombination, perhaps not by coincidence, the pressure in the Universe decreased from one-third of its density to almost zero. The earlier structures, frozen till then because of pressure, started to evolve only after that. Therefore the density (temperature) profile obtained from the image gives us the wealth of information about much earlier epochs than the plasma-transparent Universe transition. In fact, because of the above-mentioned pressure, if the cosmic neutrino background is ever measured, we expect to observe similar maps to CMB even if neutrinos were released at time about 10 trillion times earlier than photons. Therefore a question of statistical properties of the CMB is of utmost importance for our understanding the very early Universe dynamics.

The usual inflationary paradigm  with purely quantum mechanical fluctuations predicts absence of any large scale structures except the shape of the power spectrum (there can be some large structures that evolved later like domain walls. cosmic strings etc.). However, there are proposals, most notably Conformal Cyclic Cosmology (CCC) of Roger Penrose (Penrose 2010, Gurzadyan $\&$ Penrose 2013), that predict presence of large scale structures, in this case ring-type. While their work was concentrated on low-variance circles we look for differences of the temperature distributions in the inner and outer rings (to be described in detail below). Our approach should in principle ease the criticism that the observed large scale structures may be influenced heavily by the adding of the noise. 

There remains, however, a fundamental problem of the definition of statistical significance of such structures in the case of our Universe: We have only one map. The map can be viewed and analyzed from different perspectives, in different frequency bands, and so on, but the situation is dramatically different from, say, particle physics where we can in principle get as many confidence level digits as we want at the expense of additional observational effort. The (dis)proof of the presence of such structures can therefore never be fully convincing and depends in an essential way on the initial assumptions.

The problem of finding structures in CMB maps was considered in a number of papers, in particular (Gurzadyan $\&$ Penrose 2013, Bielewicz, Wandelt $\&$ Banday 2013, DeAbreu, Contreras $\&$ Scott 2015, Hajian 2011, Moss, Scott $\&$ Zibin 2011, N\ae ss 2012, Wehus $\&$ Eriksen 2011). The first paper at this list provides an evidence in favor of the existence of large structures while the others do not find the evidence convincing (but the conclusions sometimes disagree with each other). Our previous attempts, (Meissner, Nurowski $\&$ Ruszczycki 2013, An, Meissner $\&$ Nurowski 2013), were devoted to the search of such structures on the WMAP and on the first release of the Planck data (Planck Collaboration 2013). Neither the first one nor the second paper used the point source mask to eliminate the known point sources on the sky that could have influenced the result. The results of (An, Meissner $\&$ Nurowski 2013) based on (Planck Collaboration 2013) seemed to have been a little inconsistent with (Meissner, Nurowski $\&$ Ruszczycki 2013) so with the release of the new data by the Planck collaboration (Cosmological Parameters 2015) we have decided to carry out a new analysis and this paper reports the results.

In the present project we looked at ring-type structures in the real CMB temperature map in the frequency band 70 GHz as measured by the Planck collaboration as well as the foreground removed maps SMICA, SEVEM and NILC\footnote{Another foreground removed map called Commander-Ruler was also investigated, but the results from this map are not presented here because the results among the foreground removed maps were not much different.} 
Rather than looking directly for ring structures, we present a case that certain cumulative distribution functions of the real maps are distinguished from that of the simulated maps generated by the HEALPix code (G{\'o}rski et al. 2005).

Before discussing the procedure we would like to address two issues regarding comparison of the real map with the simulated maps. 

The first is the noise and foreground issue. To avoid the foreground, we took only those points whose disc of radius $0.42$ does not overlap\footnote{We use the radius $0.42$ because the width of the ring is $0.08$ and the largest radius we used was $0.34$} with the foreground map in more than $1\%$ of the area. We also took various foreground-removed maps provided by the Planck team to compare. For the noise, since our calculation involves average temperatures for large structures, it should not affect the analysis. So we have not added noise to the simulated maps nor the real map. However, we chose 70GHz map as the main choice because it has the least noise among the real maps. Also we took the power spectrum and smoothed out both the real map and the simulated map by cutting off at frequencies $\ell>1500$ because the data is noise-dominated above that frequency. 

The second issue is the power spectrum and beams. One has to make sure that we are comparing maps with the same power spectrum or the comparison would be meaningless. But presence of noise and foreground structures make it impossible to know the exact power spectrum of the real map. Therefore, we used two different power spectra for our result and cross checked our result with three additional spectra to make sure that our results do not vary too much among the power spectra. The two power spectrum used in this paper are the (binned) maximum likelihood power spectrum obtained from the 70GHz map, and one of the theoretical power spectrum\footnote{More specifically, it is the base\_plikHM\_TTTEEE\_lowTEB\_lensing\_post\_BAO power spectrum downloaded from Planck PR2 Ancillary data. The acronym is explained in (Cosmological Parameters 2015)}. We also used simulated Gaussian beams (given by $fwhm$ size in arcmins)  along with the power spectrum when generating the simulated maps.

\section{Procedure}

The following is an outline of the procedure we have taken.

\begin{enumerate}

\item
For each power spectrum described above, we generated two sets of 1000 simulated maps using the HEALPIX code with $NSIDE=2048$, $lmax=1500$. The first 1000-maps set was generated with $fwhm$=13 arcmin which is the effective beam fwhm value of the Planck 70GHz map (Planck 2014 XXVIII) and the second 1000-maps set was generated with $fwhm$=5 arcmin which is the effective beam fwhm value (Planck 2013 XII) of the foreground removed maps SMICA,SEVEM and NILC. 
\item
A set of $49152$ possible center points with HEALPix $NSIDE=64$ parametrization spreading over the entire celestial sphere has been created. Then for each of the points in the $NSIDE=64$ grid, we put a disc of radius $0.42$ radians and calculated how much the disc overlaps with the galactic mask and point source mask of the 70GHz map. The masks were obtained from Planck Ancillary Data. We chose only those centers which the overlap is less than $1\%$. This criteria gave us $14924$ centers for the rings. 

\item For the real maps, we chose the 70GHz Planck map and foreground removed maps SMICA, SEVEM, and NILC, all provided by Planck team. All maps were smoothed by cutting off frequencies above $\ell>1500$ to reduce noise, and resolution was downgraded to $NSIDE=1024$ in order to reduce the computational load. Same downgrade of the resolution was done to the simulated maps also.

\item
We were considering circles $C_{(i,\ga)}$ with a spherical radius $\gamma$ centered at the $N_d=14924$ points $(\theta^i,\phi^i)$, $i=1,2,\dots,N_d$, from the grid on the sphere. Each circle $C_{(i,\ga)}$ was surrounded by two rings - an inner ring $R_{(i,\ga,\eps_-)}$, and an outer ring $R_{(i,\ga,\eps_+)}$ - each of width $\eps$. The inner (respectively, outer) ring consisted of $N_{(i,\ga,\eps_-)}$ (respectively, $N_{(i,\ga,\eps_+)}$) points, whose spherical angle from the point $(\theta^i,\phi^i)$ was between $\ga-\eps$ and $\ga$ (respectively, between $\ga$ and $\ga+\eps$). The points in the rings excluded all the points that belong to the galactic mask or the point source mask. This exclusion was done in all computations including the computation on simulated maps, in order to have a fair comparison. 

\item For each $(\theta^i,\phi^i)$, each $\ga=0.10$, 0.14, 0.18, 0.22, 0.26, 0.30, 0.34 radians and a fixed width $\eps=0.08$ radians, we calculated the difference between the average temperature in the inner ring and outer ring, i.e. the quantity
\beq
I_{(\ga,\eps)}(\theta^i,\phi^i)=\sum_{{\rm points~in}~R_{(i,\ga,\eps_-)}}\hspace{-0.5cm}\frac{\de T_m}{N_{(i,\ga,\eps_-)}}\quad-\sum_{{\rm points~in}~R_{(i,\ga,\eps_+)}}\hspace{-0.5cm}\frac{\de T_m}{N_{(i,\ga,\eps_+)}}\nn\quad
\eeq
with $\de T_m$ being the temperatures at the points in the respective rings. Since $7$ different radii were considered for $14924$ centers, a total of $104468$ integral values were computed per map.
\item
For each $\ga$'s we calculated the standard deviation $\sigma_{\ga}$ of the integral values and defined the normalization
\beq
\hat{I}_{(\ga,\eps)}(\theta^i,\phi^i)=I_{(\ga,\eps)}(\theta^i,\phi^i)/\sigma_{\ga}.
\eeq
This normalization was necessary in order to compare integral values between maps with different power spectra.
\item
From the normalized integral values of the simulated maps, we constructed cumulative distribution function $F_{\ga}(\hat{I})$ for each radius $\ga$.
\item
For each simulated maps and the Planck maps, we computed A-values defined as:
\beq
A^+_{\ga} = -\frac{a}{N_d} \sum_{i=1}^{N_d} \ln(1-F_{\ga}(\hat{I_i})^a)
\eeq
\beq
A^-_{\ga} = -\frac{a}{N_d} \sum_{i=1}^{N_d} \ln(1-(1-F_{\ga}(\hat{I_i}))^a)
\eeq
where $a=10000$ and $I_i$ is the normalized integral values. The A-values enable us to compare the right and left tail of a CDF to another CDF (Meissner, Nurowski $\&$ Ruszczycki 2013). 
\end{enumerate}

Since we considered $7$ different radii, each map gives us $14$ A-values. The advantage of the $A^+$ value (respectively $A^-$ value) is that it is sensitive to right end (respectively left end) of the CDF because the extremal values of the integral give the biggest contribution in the summand. It also has established statistical properties (see (Meissner 2012)). Thus using these A-values we can compare the real maps and the simulated maps.

\section{Results}

Let us repeat how our procedure of comparing real maps with the simulated maps is actually performed (see (Meissner 2012) for detailed statistical theory arguments).

We simulate a large number (in our case 1000) of simulated maps with a given theoretical power spectrum and, for each map, we find a cumulative distribution function (CDF$(\gamma,\epsilon)$) for the overlap integrals of ring regions (of a given radius $\gamma$ and width $\epsilon$) pointing in many different directions on the map. After producing 1000 CDF$(\gamma,\epsilon)$s we `average' them over 1000 maps, to create a 'theoretical' CDF$(\gamma,\epsilon)$. Then, we use the theoretical CDF$(\gamma,\epsilon)$ to calculate $A$ values (right $A^+$ and left $A^-$) for each of the simulated maps and the real map. At this point, although it is desirable to apply the statistical procedure given in (Meissner 2012), the correlation of the integral values among nearby rings (i.e. for a fixed radius and width, the closer the centers of the two rings are, the closer their integral values will be) makes it very hard to do so. Therefore, we have to resort to a bit less precise but simpler quantitative estimate of the probability, namely the percentage of the simulated maps having the $A$-value higher than the $A$-value of the real map.

We confront this number with the hypothesis that both the real and the averaged CDF$(\gamma,\epsilon)$ are the same. If this hypothesis was true we would expect that on average half of the simulated maps should have the $A$-value bigger and half smaller than the corresponding $A$-value of the real map. Our results displayed in the tables 1 and 2 show that there is a distinguished value of the radius $\gamma$ were this situation does not occur, indicating that possibly the distribution of the circles with this radii on the real map is not purely statistical. 

As we can see, among the different radii, the $A^-$ value of $\ga=0.14$ gives strong significance at the level of $99\%$ or above. Most notably, the 70GHz map, when compared to simulated maps generated with the (binned) maximum likelihood power spectrum of the 70GHz map, it reaches  $99.7\%$. Further away from $\ga=0.14$ the result approaches the statistical fluctuations around 50\%.

\begin{table}
        \centering
\renewcommand{\arraystretch}{1.3}
        \caption{Significance of A values for binned power spectrum (for width $\eps=0.08$)}
        \begin{tabular}{rrrcrrcrrcrr}
        \toprule
        \phantom{abc}  & \multicolumn{2}{c}{70GHz map} & \phantom{abc} & \multicolumn{2}{c}{SMICA}  & \phantom{abc} & \multicolumn{2}{c}{SEVEM} & \phantom{abc} & \multicolumn{2}{c}{NILC} \\
                \cmidrule{2-3} \cmidrule{5-6} \cmidrule{8-9} \cmidrule{11-12}
        Radius        &$A^+$ &  $A^-$ &\phantom{abc} & $A^+$ &  $A^-$ &\phantom{abc} & $A^+$ &  $A^-$ &\phantom{abc} & $A^+$ &  $A^-$ \\
        \midrule
0.10 & 47.9\% & 57.4\% & \phantom{abc} & 42.9\% & 51.2\% & \phantom{abc} & 44.0\% & 49.2\% & \phantom{abc} & 42.0\% & 51.3\%\\ 
0.14 & 2.3\% & 99.7\% & \phantom{abc} & 2.5\% & 98.9\% & \phantom{abc} & 2.5\% & 98.9\% & \phantom{abc} & 3.7\% & 98.9\%\\ 
0.18 & 30.3\% & 80.2\% & \phantom{abc} & 33.3\% & 76.2\% & \phantom{abc} & 30.9\% & 77.7\% & \phantom{abc} & 33.0\% & 78.5\%\\ 
0.22 & 31.2\% & 21.0\% & \phantom{abc} & 41.5\% & 12.6\% & \phantom{abc} & 40.6\% & 17.9\% & \phantom{abc} & 40.0\% & 16.6\%\\ 
0.26 & 75.7\% & 49.2\% & \phantom{abc} & 86.1\% & 38.8\% & \phantom{abc} & 85.2\% & 45.9\% & \phantom{abc} & 84.8\% & 43.8\%\\ 
0.30 & 42.6\% & 43.1\% & \phantom{abc} & 50.6\% & 46.1\% & \phantom{abc} & 46.4\% & 45.9\% & \phantom{abc} & 45.9\% & 47.5\%\\ 
0.34 & 45.6\% & 94.4\% & \phantom{abc} & 55.0\% & 93.8\% & \phantom{abc} & 51.5\% & 93.1\% & \phantom{abc} & 49.9\% & 93.9\%\\ 
    \end{tabular}
\end{table}

\begin{table}
        \centering
\renewcommand{\arraystretch}{1.3}
        \caption{Significance of A values for theoretical power spectrum (for width $\eps=0.08$)}
        \begin{tabular}{rrrcrrcrrcrr}
        \toprule
        \phantom{abc}  & \multicolumn{2}{c}{70GHz map} & \phantom{abc} & \multicolumn{2}{c}{SMICA}  & \phantom{abc} & \multicolumn{2}{c}{SEVEM} & \phantom{abc} & \multicolumn{2}{c}{NILC} \\
                \cmidrule{2-3} \cmidrule{5-6} \cmidrule{8-9} \cmidrule{11-12}
        Radius        &$A^+$ &  $A^-$ &\phantom{abc} & $A^+$ &  $A^-$ &\phantom{abc} & $A^+$ &  $A^-$ &\phantom{abc} & $A^+$ &  $A^-$ \\
        \midrule
0.10 &46.4\% & 57.7\% & \phantom{abc} & 45.4\% & 54.7\% & \phantom{abc} & 46.3\% & 51.8\% & \phantom{abc} & 44.7\% & 55.5\%\\ 
0.14 &2.6\% & 99.2\% & \phantom{abc} & 3.2\% & 99.1\% & \phantom{abc} & 3.1\% & 99.1\% & \phantom{abc} & 4.6\% & 99.1\%\\ 
0.18 &31.7\% & 73.3\% & \phantom{abc} & 33.2\% & 72.6\% & \phantom{abc} & 30.2\% & 74.6\% & \phantom{abc} & 33.0\% & 75.8\%\\ 
0.22 &33.4\% & 19.8\% & \phantom{abc} & 42.5\% & 13.5\% & \phantom{abc} & 41.4\% & 19.5\% & \phantom{abc} & 40.2\% & 17.8\%\\ 
0.26 &74.8\% & 50.6\% & \phantom{abc} & 86.5\% & 41.4\% & \phantom{abc} & 85.7\% & 47.8\% & \phantom{abc} & 85.3\% & 45.8\%\\ 
0.30 &45.1\% & 43.0\% & \phantom{abc} & 49.2\% & 48.4\% & \phantom{abc} & 46.3\% & 48.0\% & \phantom{abc} & 45.4\% & 50.6\%\\ 
0.34 &44.1\% & 94.9\% & \phantom{abc} & 54.9\% & 94.1\% & \phantom{abc} & 53.0\% & 93.7\% & \phantom{abc} & 51.3\% & 94.1\%\\  
    \end{tabular}
\end{table}

Since this radius $\gamma=0.14$ radians produced numbers that are very different than the rest, we looked at various radii and widths to see if there is an even more significant case. The result is shown in table 3, where we computed and compared the $A$ values for the radii $\ga=0.12,0.13,0.14,0.15$ and $0.16$, and widths $\eps=0.06,0.07,0.08$ and $0.09$. We see that for radius $\ga=0.14$, the width $\eps=0.06$ produced $99.9\%$, which means it produced higher $A$ value than 999 simulated maps of the 1000 simulated maps.    

\begin{table}
        \centering
\renewcommand{\arraystretch}{1.3}
        \caption{Significance of A values of 70GHz map near $\gamma=0.14$ and $\eps=0.08$}
        \begin{tabular}{|c|cr|cr|cr|cr|}
        \hline
        \diaghead{\theadfont radius n width}%
{radius}{width} &\phantom{abc}&0.06 &\phantom{abc}&  0.07 &\phantom{abc}& 0.08 &\phantom{abc}& 0.09 \\
        \hline
0.12 &\phantom{abc}& 90.2\% &\phantom{abc}& 96.9\% &\phantom{abc}& 98.3\% &\phantom{abc}& 98.6\%\\ 
0.13 &\phantom{abc}& 99.2\% &\phantom{abc}& 99.5\% &\phantom{abc}& 99.5\% &\phantom{abc}& 99.3\%\\ 
0.14 &\phantom{abc}& 99.9\% &\phantom{abc}& 99.8\% &\phantom{abc}& 99.7\% &\phantom{abc}& 99.4\%\\ 
0.15 &\phantom{abc}& 99.3\% &\phantom{abc}& 99.1\% &\phantom{abc}& 98.8\% &\phantom{abc}& 98.9\%\\ 
0.16 &\phantom{abc}& 92.6\% &\phantom{abc}& 93.6\% &\phantom{abc}& 94.8\% &\phantom{abc}& 95.9\%\\ 
\hline
    \end{tabular}
\end{table}

We also checked whether circles are distinguished when compared with deformed ellipses. We use the deforming method proposed by R. Penrose to twist the southern and northern hemispheres (i.e. we search for deformed ellipses). It is important to note that while this procedure would change the power spectrum, the simulated maps are also twisted using the same procedure, and therefore the comparison is still being done in a fair way. We show below the number of simulated maps with the $A$ value exceeding the real map for different twistings  (again for $\ga=0.14$ radians).  We see that they consistently and rapidly approach a purely statistical result with the growth of twisting :\\

\begin{center}
\resizebox{7cm}{4cm}{%
\begin{tikzpicture}
\begin{axis}[
    title={},
    xlabel={twisting},
    ylabel={number of maps},
    xmin=0, xmax=10,
    ymin=0, ymax=650,
    xtick={0,1,2,3,4,5,6,7,8,9},
    ytick={0,50,100,150,200,250,300,350,400,450,500,550,600},
    legend pos=north west,
    ymajorgrids=true,
    grid style=dashed,
]

\addplot[
    color=red,
    mark=triangle,
    ]
    coordinates {
    (0,11)(1,10)(2,16)(3,32)(4,79)(5,140)(6,234)(7,353)(8,510)(9,586)
    };

\end{axis}
\end{tikzpicture}
}
\end{center}

Figs. 1 and 2 show the rings which gave a high integral value, whose significance is judged by using the cumulative distribution functions obtained from the simulated maps generated from binned power spectrum.

\begin{figure}[htb]
\begin{center}
\includegraphics[scale=0.32]{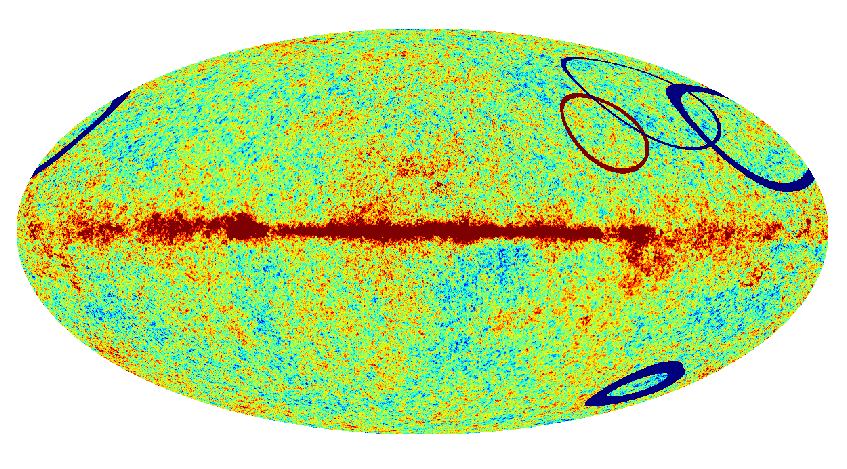}
\caption{\scriptsize{Picture of 25 rings that have significance less than .01\% (marked blue) or more than 95.99\% (marked red) when measured using the CDF obtained from the simulated maps. The  rings investigated here have radii $\gamma$=[0.10,0.14,0.18,0.22,0.16,0.30,0.34] and width $\eps$=0.08 radians. The blue circles have higher temperature on the outer ring than the inner ring and red vice versa. The rings are drawn on the 70GHz Planck map. }}
\end{center}
\end{figure}

\begin{figure}[htb]
\begin{center}
\includegraphics[scale=0.32]{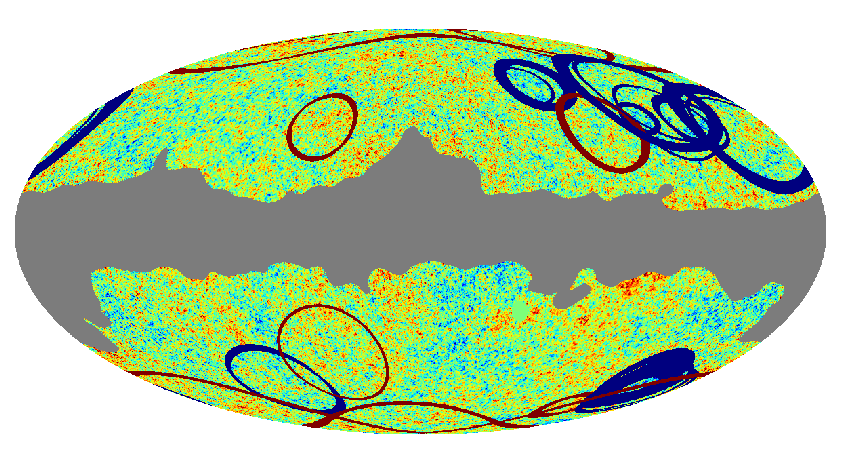}
\caption{\scriptsize{Picture of 115 rings that have significance less than .05\% (marked blue) or more than 95.95\% (marked red) when measured using the CDF obtained from the simulated maps. The  rings investigated here have radii $\gamma$=[0.10,0.14,0.18,0.22,0.16,0.30,0.34] and width $\eps$=0.08 radians. This figure also includes the mask used for the computation.}}
\end{center}
\end{figure}

\begin{figure}[htb]
\begin{center}
\includegraphics[scale=0.45]{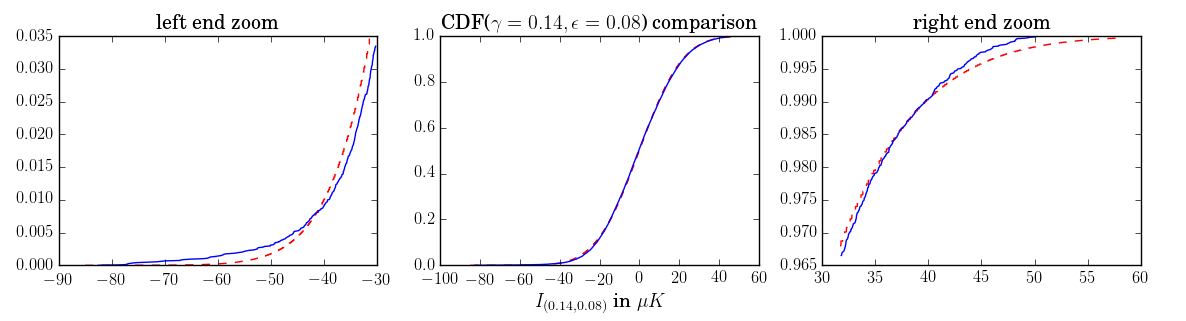}
\caption{\scriptsize{The dotted curve is the CDF(0.14,0.08) obtained from the simulated maps. The solid curve is the cumulative distribution of the integral values on the 70GHz map. As you can see in the middle graph, they agree quite well in general. The lower end and the higher end of the CDF are zoomed in to see the comparison of the two at the extremes, showing that the distribution of the real map deviates slightly from the CDF of simulated maps at both ends. }}
\end{center}
\end{figure}

\section{Conclusions}
The results presented show that there is a distinguished radius $\ga=0.14$ for which the number of simulated maps with lower $A$-value than the real map is very high (around 99.7\%) while for both higher and lower radii the number rapidly goes down. An attempt to look for deformed ellipses by twisting gives also a rapid growth of the simulated maps with bigger $A$-value than the twisted real map. Both these results suggest that the circular structures in the vicinity of the radius $\ga=0.14$ radians may have a non-statistical origin.

\vspace{3mm}

\noindent{\bf {Acknowledgments:}} We gratefully acknowledge helpful discussions with Pawe{\l} Bielewicz, Marek Demia\'nski, Sigurd K. N{\ae}ss, C. Denson Hill, George Efstathiou, E. Ted Newman, Roger Penrose, Vahe Gurzadyan and  B{\l}a{\.z}ej Ruszczycki. KAM was supported by the Polish NCN grant DEC-2013/11/B/ST2/04046, and PN by DEC-2013/09/B/ST1/01799.

\vspace{3mm}
\centerline{REFERENCES}

\noindent
An D., Meissner K.A. and Nurowski P., 2013, {\it Structures in the Planck map of the CMB}, eprint arXiv:1307.5737\\[1mm]
Bielewicz P., Wandelt B.D. and Banday A.J., 2013, {\it A search for concentric rings with unusual variance in the 7-year WMAP temperature maps using a fast convolution approach}, Monthly Notices of the Royal Astronomical Society {\bf 429}, 1376\\[1mm]
DeAbreu A., Contreras D. and Scott D., 2015, {\it Searching for concentric low variance circles in the cosmic microwave background}, eprint arXiv:1508.05158\\[1mm]
G\'orski K.M. et al., 2005, {\it HEALPix: A Framework for High-Resolution Discretization and Fast Analysis of Data Distributed on the Sphere}, Astroph. J. {\bf 622} 759-771.\\[1mm]
Gurzadyan V.G. and Penrose R., 2013, {\it On CCC-predicted
concentric low-variance circles in CMB sky.} Eur. Phys. J. Plus {\bf 128}, 22\\[1mm]
Hajian A., 2011, {\it Are there Echoes from the Pre-big-bang Universe? A Search for Low-variance Circles in the Cosmic Microwave Background Sky}, The Astrophysical Journal {\bf 740}, 52\\[1mm]
Meissner K. A., 2012, {\it A Tail Sensitive Test for Cumulative Distribution Functions}, {\tt arXiv:1206.4000 [math-st]}\\[1mm]
Meissner K.A., Nurowski P. and Ruszczycki B., 2013, {\it Structures in the microwave background radiation}, Proc. R. Soc. {\bf A469}:2155, 20130116\\[1mm]
Moss A., Scott D. and Zibin J.P., 2011, {\it No evidence for anomalously low variance circles on the sky}, Journal of Cosmology and Astroparticle Physics 4, 33\\[1mm]
N\ae ss S.K., 2012, {\it Application of the Kolmogorov-Smirnov test to CMB data: Is the universe really weakly random?}, Astronomy \& Astrophysics {\bf 538}, A17\\[1mm]
Penrose R., 2010, {\it Cycles of Time: An Extraordinary New View of the Universe}, Bodley Head, \\[1mm]
Planck Collaboration, 2013, {\it Planck 2013 results}, http://planck.caltech.edu/publications2013Results. html\\[1mm]
Planck Collaboration, 2013 {\it Planck 2013 results, XII. Component separation}\\[1mm]
Planck Collaboration, 2014 {\it Planck Collaboration XXVIII, 2014 } A\&A, 571, A28\\[1mm]
Planck Collaboration, 2015 {\it Planck 2015 results. XIII. Cosmological parameters} arXiv:1502.01589 
Wehus I.K. and Eriksen H.K., 2011, {\it A Search for Concentric Circles in the 7 Year Wilkinson Microwave Anisotropy Probe Temperature Sky Maps}, The Astrophysical Journal Letters {\bf 733}, L29\\[1mm]
2015 Cosmological parameters and MC chains, http://wiki.cosmos.esa.int/planckpla2015/index. php/Cosmological\_Parameters\\[1mm]

\end{document}